\documentclass[10pt,twocolumn,prl,aps,amssymb,amsmath,tightenlines,showpacs]{revtex4}

\usepackage{graphicx}

\newcommand{\cC}{\ensuremath{\mathcal{C}}}

\newcommand{\cP}{\ensuremath{\mathcal{P}}}
\newcommand{\cT}{\ensuremath{\mathcal{T}}}
\newcommand{\cD}{\ensuremath{\mathcal{D}}}

\begin{document}

\title{Families of particles with different masses in PT-symmetric quantum field theory}

\author{Carl M. Bender$^a$}\email{cmb@wustl.edu}
\author{S. P. Klevansky$^b$}\email{spk@physik.uni-heidelberg.de}
\affiliation{$^a$Physics Department, Washington University, St. Louis, MO 63130,
USA}
\affiliation{$^b$Institut f\"ur Theoretische Physik, Universit\"at Heidelberg,
Philosophenweg 19, 69120 Heidelberg, Germany}

\date{\today}

\begin{abstract}
An elementary field-theoretic mechanism is proposed that allows one Lagrangian
to describe a family of particles having different masses but otherwise similar
physical properties. The mechanism relies on the observation that the
Dyson-Schwinger equations derived from a Lagrangian can have many different but
equally valid solutions. Nonunique solutions to the Dyson-Schwinger equations
arise when the functional integral for the Green's functions of the quantum
field theory converges in different pairs of Stokes' wedges in complex field
space, and the solutions are physically viable if the pairs of Stokes' wedges
are $\cP\cT$ symmetric.
\end{abstract}

\pacs{11.30.Er, 03.65.Db, 11.10.Ef}

\maketitle

The standard model of elementary particles has three generations of fermions
(leptons and quarks) whose masses range over several orders of magnitude. It is
not known why there are three generations of masses and whether there are only
three. This paper proposes a field-theoretic mechanism that might explain the
occurrence of generations of particles having different masses but otherwise
similar physical properties: There might be just one Lagrangian (or Hamiltonian)
to account for the properties of all these particles, but the functional
integral constructed from this Lagrangian may have many different physical
realizations depending on the boundary conditions on the path of integration in
complex field space. While the Dyson-Schwinger equations constructed from the
functional integral are unique, the solution to these equations is not unique.
The number of distinct solutions to the Dyson-Schwinger equations equals the
number of pairs of complex Stokes' wedges in function space in which the
boundary conditions on the functional integration can be imposed. For each pair
of Stokes' wedges there corresponds a different field theory.

Z.~Guralnik {\it et al} \cite{R1} first recognized that for functional
integrals, inequivalent classes of contours associated with different complex
boundary conditions give rise to nonunique solutions to the Dyson-Schwinger
equations. They argued that multiple solutions might account for inequivalent
$\theta$ vacua. The key point of the current paper is that the pairs of Stokes'
wedges in which the integration contours terminate must be oriented in a $\cP
\cT$-symmetric fashion in complex field space. If this is the case, there is
strong evidence that the corresponding field theory will be physically
acceptable; that is, the masses (poles of the Green's functions) will be real
and the theory will be unitary. The mechanism proposed here is field-theoretic,
but its application is not restricted to elementary particle physics.
Experiments on $\cP \cT$-symmetric optical wave guides \cite{R2,R3} and on $\cP
\cT$-symmetric diffusion \cite{R4} have been reported recently.  

The conjecture discussed in this paper stems from recent research on $\cP\cT$
quantum mechanics, where it has been shown that the $\cP\cT$-symmetric
Hamiltonians
\begin{equation}
H=p^2+q^2(iq)^\epsilon\quad(\epsilon\geq0)
\label{e1}
\end{equation}
all have real positive spectra \cite{R5,R6}. Each of these Hamiltonians defines
a conventional quantum theory with a Hilbert space having a positive inner
product \cite{R7}. The time-evolution operator $U=e^{-iHt}$ is unitary and thus
probability is conserved. Spectral reality and unitary time evolution are
essential for any quantum theory. These features are guaranteed if $H$ is Dirac
Hermitian. (By {\it Dirac Hermitian} we mean that $H=H^\dag$, where $\dag$
represents combined complex conjugation and matrix transposition.) However, it
is not necessary for $H$ to be Dirac Hermitian for the spectrum to be real and
for time evolution to be unitary; non-Dirac-Hermitian Hamiltonians can also
define physically acceptable quantum theories.

The Hamiltonians (\ref{e1}) are $\cP\cT$ symmetric because they are invariant
under combined spatial reflection $\cP$ and time reversal $\cT$. Such
Hamiltonians are physically acceptable because they are selfadjoint, not with
respect to the Dirac adjoint $\dag$, but rather with respect to $\cC\cP\cT$
conjugation, where $\cC$ is a linear operator that represents a hidden
reflection symmetry of $H$. The $\cC\cP\cT$ adjoint defines a positive-definite
Hilbert space norm. Not every $\cP\cT$-symmetric Hamiltonian has an entirely
real spectrum, but the spectrum {\it is} entirely real if and only if a linear
$\cP\cT$-symmetric operator $\cC$ exists that obeys three simultaneous
algebraic equations \cite{R7}: $\cC^2=1$, $[\cC,\cP\cT]=0$, $[\cC,H]=0$. When
the $\cC$ operator exists, we say that the $\cP\cT$ symmetry of $H$ is {\it
unbroken}. Finding the $\cC$ operator is the crucial step in showing that time
evolution for a non-Hermitian $\cP\cT$-symmetric Hamiltonian is unitary. The
phase transition between broken and unbroken regions for some $\cP\cT$-symmetric
Hamiltonians has been observed experimentally \cite{R3,R4}.

The Hamiltonians in (\ref{e1}) are smooth extensions in the parameter $\epsilon$
of the Dirac-Hermitian harmonic oscillator Hamiltonian (at $\epsilon=0$) into
the complex non-Hermitian domain ($\epsilon>0$). As $\epsilon$ increases from 0,
the Stokes' wedges in the complex-$x$ plane inside of which the boundary
conditions for the eigenvalue problem
\begin{equation}
-\psi^{\prime\prime}(x)+x^2(ix)^\epsilon\psi(x)=E\psi(x)
\label{e2}
\end{equation}
are imposed, rotate downward and become thinner. As shown in Ref.~\cite{R5}, at
$\epsilon=0$ the Stokes' wedges are centered about the positive- and
negative-real axes and have angular opening $90^\circ$. At $\epsilon=2$
the Stokes' wedges are adjacent to and below the real axes and have angular
opening $60^\circ$. When $\epsilon>2$, these wedges lie below the real axis.

To illustrate the idea of this paper in a quantum-mechanical context we set
$\epsilon=4$ in (\ref{e1}). The resulting $x^6$ Hamiltonian describes {\it two}
different quantum theories because the eigenfunctions $\psi(x)$ can satisfy two
different sets of boundary conditions \cite{R8}: (i) the conventional
Dirac-Hermitian quantum theory for which $\psi(x)$ vanishes as $|x|\to\infty$ in
the complex-$x$ plane in $45^\circ$ wedges centered about the real axes; or (ii)
the unconventional $\cP\cT$ theory, which is the extension in $\epsilon$ of the
harmonic oscillator. For this non-Hermitian quantum theory $\psi(x)$ also
vanishes as $|x|\to\infty$ in the complex plane in $45^\circ$ wedges, but now
these wedges are centered about ${\rm arg}\,x=-45^\circ$ and ${\rm arg}\,x=
-135^\circ$. The one-point Green's function $G_1=\langle x\rangle$ distinguishes
between these two theories. The conventional Dirac-Hermitian theory has parity
symmetry, and thus $G_1$ vanishes. The boundary conditions for the $\cP\cT$
quantum theory violate parity symmetry, and as a result $G_1$ has a
negative-imaginary value. The nonvanishing of $G_1$ in the $\cP\cT$ theory is a
purely nonperturbative effect; one cannot express $G_1$ for the Hamiltonian
$H=p^2+x^2+gx^6$ as a series in powers of $g$.

% CCCCCCCCCCCCC

The idea that different boundary conditions allow one Hamiltonian (or
Lagrangian) to describe several different physical theories is general and
extends beyond quantum mechanics to quantum field theories of fermion and/or
boson fields of any spin and in any space-time dimension. However, for brevity
we consider here the massless $D$-dimensional pseudoscalar field theory ($D<2$)
having a selfinteraction of the form $\phi^{4n+2}$ ($n=1,\,2,\,3,\,\ldots$).
(Under parity reflection $\phi\to-\phi$.) The Euclidean functional integral for
the vacuum persistence functional in the presence of an external source $J$ is
\begin{eqnarray}
Z[J]&=&\langle0|0\rangle=\int_C\cD\phi\,e^{-S},\nonumber\\
S&=&\int d^Ds\left[\frac{1}{2}(\nabla\phi)^2+\frac{g}{4n+2}\phi^{4n+2}-J\phi
\right].
\label{e3}
\end{eqnarray}
At $D=1$, this quantum field theory reduces to a quasi-exactly-solvable
quantum-mechanical theory \cite{R9}.

\begin{figure}[h!]
\begin{center}
\includegraphics[scale=0.55, bb=0 0 347 285]{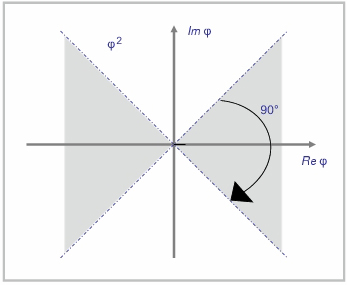}
\end{center}
\caption{Stokes' wedges (shaded regions) in the complex-$\varphi$ plane in which
the integration contour $C$ for the integral $\int_C d\varphi\,\exp\left(-
\varphi^2\right)$ terminates. This integral does not exist if $C$ terminates in
an unshaded wedge.}
\label{F1}
\end{figure}

\begin{figure}[h!]
\begin{center}
\includegraphics[scale=0.55, bb=0 0 348 284]{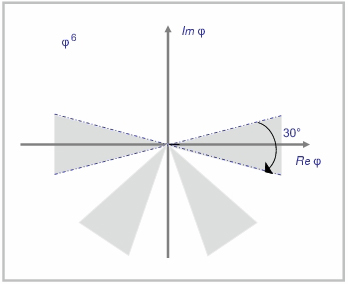}
\end{center}
\caption{Stokes' wedges (shaded regions) of angular opening $30^\circ$ in which
the integration contour $C$ for the integral $\int_C d\varphi\,\exp\left(-
\varphi^6\right)$ may terminate. The integral has two possible real values, one
for which the contour connects the pair of wedges centered about the real axis
and the other for which the contour connects the lower pair of wedges.}
\label{F2}
\end{figure}

For each integer $n$ there are $n+1$ different physical realizations of the
quantum field theory in (\ref{e3}). To explain this we consider the analogous
one-dimensional integral $\int_C d\varphi\,\exp\left(-\varphi^{4n+2}\right)$.
When $n=0$ this integral exists only if the integration contour $C$ begins and
ends in the Stokes' wedges of angular opening $90^\circ$ centered about the
real-$\varphi$ axis. These Stokes' wedges are shown in Fig.~\ref{F1}. The
contour $C$ must begin and end in different Stokes' wedges; if $C$ begins and
ends in the same Stokes' wedge, the integral vanishes. When $n=1$, there are
{\it two} possible choices for integration contour $C$; $C$ may connect the two
$30^\circ$-Stokes' wedges centered about the real axis or $C$ may connect the
$30^\circ$-Stokes' wedge centered about $-120^\circ$ to the $30^\circ$-Stokes'
wedge centered about $-60^\circ$ (see Fig.~\ref{F2}).

The contour $C$ for $\int_C d\varphi\,\exp\left(-\varphi^6\right)$ must join a
pair of $\cP\cT$-symmetric Stokes' wedges (wedges that are sym-
\begin{widetext}
\begin{figure}[h!]
\begin{center}
\includegraphics[scale=0.55, bb=0 0 685 282]{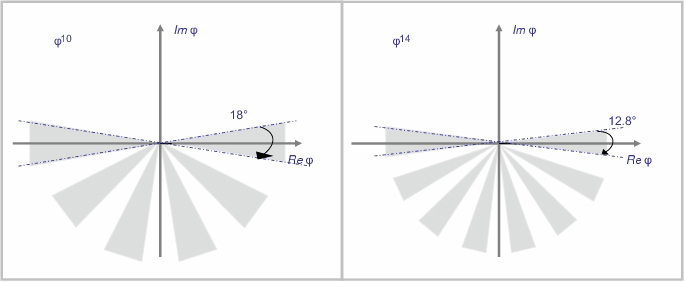}
\end{center}
\caption{Stokes' wedges (shaded regions) in which the integration contour $C$
for $\int_C d\varphi\,\exp\left(-\varphi^{4n+2}\right)$ may terminate. When
$n=2$ (left figure) there are three pairs of Stokes' wedges and when $n=4$ there
are four pairs of Stokes' wedges.}
\label{F3}
\end{figure}
\end{widetext}
metric about the imaginary axis) or else the integral is not real. A third pair
of
$30^\circ$-Stokes' wedges, one centered about $60^\circ$ and the other centered
about $120^\circ$, are not shown in Fig.~\ref{F2}; the integral exists if the
contour $C$ connects this pair of Stokes' wedges, but this case is not new;
it is just the complex conjugate of the configuration in which $C$
connects the $-120^\circ$ and $-60^\circ$ wedges.

The cases $n=2$ (three pairs of $18^\circ$ Stokes' wedges) and $n=3$ (four pairs
of $12.8^\circ$ Stokes' wedges) are shown in Fig.~\ref{F3}. In the former case
the $\int_C d\varphi\,\exp\left(-\varphi^{4n+2}\right)$ has three independent
real values; in the latter case it has four independent real values.

Returning to the quantum field theory with vacuum persistence function given in
(\ref{e3}), we vary the action in the exponent and obtain the Euclidean field
equation in the presence of the external $c$-number source $J(x)$:
\begin{equation}
-\nabla^2\phi(x)+g[\phi(x)]^{4n+1}=J(x).
\label{e4}
\end{equation}
This field equation is unique; it does not depend on the choice of complex
contour $C$.

The expectation value of (\ref{e4}) in the vacuum state is
\begin{equation}
-\nabla^2G_1(x)+g\langle[\phi(x)]^{4n+1}\rangle/Z[J]=J(x),
\label{e5}
\end{equation}
where $G_1(x)$ is the connected one-point Green's function:
\begin{equation}
G_1(x)=\frac{\delta\ln Z[J]}{\delta J(x)}=\frac{\langle\phi(x)\rangle}{Z[J]}
=\int_C\cD\phi\,\phi(x)e^{-S}.
\label{e6}
\end{equation}
This expectation value depends on the choice of metric, but in Ref.~\cite{R10}
it is shown that the path integral automatically gives the expectation value
with the appropriate metric. Thus, if the integration contour terminates in the
wedges containing the real axis, then the metric uses the conventional Dirac
adjoint $\dag$, and if the contour terminates in another pair of Stokes' wedges,
then the metric uses the $\cC\cP\cT$ adjoint of the corresponding
non-Dirac-Hermitian $\cP\cT$-symmetric field theory \cite{R11}.

To derive the Dyson-Schwinger equations for the connected Green's functions of
the quantum field theory, we express the second term on the left side of
(\ref{e5}) in terms of the higher connected Green's functions. The technique
is standard (see, for example, Ref.~\cite{R12}); one differentiates repeatedly
with respect to the external source $J(x)$ and uses the formula for the
$n$-point Green's function in the presence of the external source $J$:
\begin{equation}
G_n(x,y,z,\ldots)\equiv\delta^n/[\delta J(x)\delta J(y)\delta
J(z)\cdots]\ln Z[J].
\label{e7}
\end{equation}

We must truncate the Dyson-Schwinger equations in order to obtain a closed
system. We consider here just the first two equations and neglect contributions
from Green's functions beyond $G_2(x,y)$. This truncation gives the {\it
mean-field} (or {\it one-pole}) approximation to the two-point Green's function.
(Including higher Green's functions does not change any qualitative conclusions
of this paper.) Thus, we repeatedly differentiate with respect to $J(x)$ and get
the following sequence of equations:
\begin{eqnarray}
\langle 1\rangle&=&Z[J],\quad \langle\phi(x)\rangle=G_1(x)Z[J],\nonumber\\
\langle[\phi(x)]^2\rangle&=&\left([G_1(x)]^2+G_2(x,x)\right)Z[J],\nonumber\\
\langle[\phi(x)]^3\rangle&=&\left([G_1(x)]^3+3G_1(x)G_2(x,x)\right)Z[J],
\nonumber\\
\langle[\phi(x)]^4\rangle&=&\left([G_1(x)]^4+6[G_1(x)]^2G_2(x,x)\right.
\nonumber\\
&&\left.\quad+3[G_2(x,x)]^2\right)Z[J],\nonumber\\
\langle[\phi(x)]^5\rangle&=&\left([G_1(x)]^5+10[G_1(x)]^3G_2(x,x)\right.
\nonumber\\
&&\quad\left.+15G_1[G_2(x,x)]^2\right)Z[J].
\label{e8}
\end{eqnarray}
These expressions have a simple form as polynomials $P_n(t)$ in the variable
$t=G_1(x)/\sqrt{G_2(x,x)}$,
\begin{equation}
\langle[\phi(x)]^n\rangle=[G_2(x,x)]^{n/2}Z[J]P_n(t),
\label{e9}
\end{equation}
where $P_n(t)=(-i)^n{\rm He}_n(it)$ are Hermite polynomials of imaginary
argument: $P_0(t)=1$, $P_1(t)=t$, $P_2(t)=t^2+1$, $P_3(t)=t^3+3t$, $P_4(t)=t^4
+6t^2+3$, $P_5(t)=t^5+10t^3+15t$.

Next, we insert (\ref{e9}) into (\ref{e5}) and obtain
\begin{equation}
-\nabla^2G_1(x)-i[G_2(x,x)]^{2n+1/2}{\rm He}_{4n+1}(it)=J(x).
\label{e10}
\end{equation}
At $J\equiv0$ translation invariance is restored, and $G_1(x)$ and $G_2(x,x)$
become the numbers $G_1$ and $G_2(0)$. Thus, the first of the truncated
Dyson-Schwinger equations is
\begin{equation}
{\rm He}_{4n+1}\left[iG_1/\sqrt{G_2(0)}\right]=0.
\label{e11}
\end{equation}
Note that the argument of ${\rm He}_{4n+1}$ remains invariant if wave-function
renormalization is performed.

To obtain the second Dyson-Schwinger equation we differentiate (\ref{e10}) with
respect to $J(y)$ and set $J\equiv0$:
\begin{equation}
\left(-\nabla^2+M^2\right)G_2(x-y)=\delta^D(x-y),
\label{e12}
\end{equation}
where the renormalized mass is given by
\begin{equation}
M^2=[G_2(0)]^{2n}{\rm He}_{4n+1}^\prime\left[iG_1/\sqrt{G_2(0)}\right].
\label{e13}
\end{equation}

We solve (\ref{e11})--(\ref{e13}) simultaneously: First, we Fourier transform
(\ref{e12}) and find that in $D$-dimensional Euclidean space $\tilde{G}_2(p)=
1/(p^2+M^2)$. Thus, for $0\leq D<2$ we get the finite result $G_2(0)=M^{D-2}
\Gamma(1-D/2)2^{-D}\pi^{-D/2}$. Second, we note that the Hermite polynomial
${\rm He}_{4n+1}$ is {\it odd} and only has real roots. There are two cases:
Either (i) $G_1=0$, which is the conventional Dirac-Hermitian parity-invariant
solution to the Dyson-Schwinger equations, or (ii) we get $4n$ new
parity-violating nonzero values for the one-point Green's function:
\begin{equation}
G_{1,j}=\pm iM^{-1+D/2}\sqrt{\Gamma(1-D/2)}2^{-D/2}\pi^{-D/4}r_j,
\label{e14}
\end{equation}
where the dimensionless number $r_j$ ($j=1,\ldots 2n$) is one of the $2n$
positive roots of ${\rm He}_{4n+1}$. Finally, we use the identity ${\rm He}_{4n
+1}^\prime=(4n+1){\rm He}_{4n}$ in (\ref{e13}) and use the interlacing-of-zeros
property of the Hermite polynomials to prove that there are exactly $n$ new
positive values of $M^2$ corresponding to the nonzero values of $G_{1,j}$. This
demonstrates the connection between pairs of Stokes' wedges and solutions to the
Dyson-Schwinger equations.

For example, when $D=1$ in a $\phi^6$ model, $r_0=0$ and $r_1=2.85697$, and
corresponding to these roots the dimensionless renormalized masses are
$M=1.39158$ and $M=2.25399$. Thus, there are two families of particles: One
particle (associated with a nonvanishing $G_{1,1}$) has a mass $1.62$ times
larger than that of the other particle (associated with a vanishing $G_{1,0}$). 
This ratio increases rapidly as a function of the space-time dimension $D$; for
example, for $D=0.0,~0.5,~1.0,~1.5,~2.0,~2.5$ this ratio takes the values
$1.38,~1.47,~1.62,~1.90,~2.62,~6.88$. 

To conclude, while a flavor symmetry group is conventionally introduced to
describe families of particles, we have shown that such families can arise
naturally from the monodromy structure in the complex-field plane associated
with rotation from one Stokes' wedge to another.

CMB thanks the Graduate School at the University of Heidelberg for its
hospitality and the U.S.~Department of Energy for financial support.

\end{document}